\documentclass[prd,twocolumn,showpacs]{revtex4}
\usepackage{graphicx}
\usepackage{latexsym}
\usepackage{amsmath}
\input psfig.sty

\newcommand{\bq}{\begin{equation}}
\newcommand{\eq}{\end{equation}}
\newcommand{\bqn}{\begin{eqnarray}}
\newcommand{\eqn}{\end{eqnarray}}

\begin{document}
\title{Current constraints on interacting holographic dark energy}
\author{Qiang Wu$^1$}
\email{qiang_wu@baylor.edu}
\author{Yungui Gong$^1$}
\email{yungui_gong@baylor.edu}
\author{Anzhong Wang$^1$}
\email{anzhong_wang@baylor.edu}
\author{J. S. Alcaniz$^2$}
\email{alcaniz@on.br}
\affiliation{$^1$CASPER, Physics Department, Baylor University, Waco, TX 76798, USA}
\affiliation{$^2$Observat\'{o}rio Nacional, 20921-400, Rio de Janeiro - RJ, Brasil}

\date{\today}

\begin{abstract}
Although there is mounting observational evidence that the cosmic 
expansion is undergoing a late-time acceleration, the physical 
mechanism behind such a phenomenon is yet unknown. In this paper, 
we investigate a holographic dark energy (HDE) model with interaction 
between the components of the dark sector in the light of current 
cosmological observations. We use both the new \emph{gold} sample 
of 182 type Ia supernovae (SNe Ia) and the 192 SNe Ia ESSENCE data, 
the baryon acoustic oscillation measurement from the Sloan Digital 
Sky Survey and the shift parameter from the three-year Wilkinson 
Microwave Anisotropy Probe data. In agreement with previous results, 
we show that these observations suggest a very weak coupling or even 
a noninteracting HDE. The phantom crossing behavior in the context 
of these scenarios is also briefly discussed.
\end{abstract}
\maketitle

\section{Introduction}
The current idea of a \emph{negative-pressure} dominated universe seems to be inevitable in light of the impressive convergence of the recent observational results (see, e.g., \cite{acc1,gold182,essence,sdss6,wmap3}). This in turn has led cosmologists to hypothesize on the possible existence of an exotic dark component that not only would explain these experimental data but also would reconcile them with the inflationary flatness prediction ($\Omega_{\rm{Total}} = 1$). This extra component, or rather, its gravitational effects is thought of as the first observational piece of evidence for new physics beyond the domain of the standard model of particle physics and constitutes a link between cosmological observations and a fundamental theory of nature (for some dark energy models, see \cite{ref1}. For recent reviews, see also \cite{rev}).

On the other hand, based on the effective local quantum field theories, the authors of Ref.~\cite{cohen} proposed a relationship between the ultraviolet (UV) and the infrared (IR) cutoffs due to the limit set by the formation of a black hole (BH). The UV-IR relationship in turn gives an upper bound on the zero point energy density $\rho_\Lambda\le M_p^2 L^{-2}$, which means that the maximum entropy is of the order of $S_{BH}^{3/4}$.  This zero point energy density has the same order of magnitude as the dark energy density \cite{Hsu}, and is widely referred to as the holographic dark energy (HDE) \cite{Li} (see also \cite{pad}). However, the HDE model based on the Hubble scale as the IR cutoff seems not to be able of leading to an accelerating universe \cite{Hsu}. A solution to this matter was subsequently given in Ref.~\cite{Li} that discussed the possibilities of the particle and the event horizons as the IR cutoff, and found that only the event horizon identified as the IR cutoff gave a viable HDE model \cite{Li}. The HDE model based on the event horizon as the IR cutoff was found to be consistent with the observational data \cite{gong}.

A subsequent development concerning the idea of a holographic dark energy is the possibility of considering interaction between this latter component and the dark matter in the context of a holographic dark energy model with the event horizon as the IR cutoff.  As an interesting result, it was shown that the interacting HDE model realized the phantom crossing behavior \cite{intde}, which is also obtained in the context of non-minimally coupled scalar fields (see, e.g, \cite{peri} and references therein). Other recent discussions on interacting HDE models can be found in \cite{holo,pavon,kim}.

In this paper, we test the viability of the interacting HDE model discussed in Ref.~\cite{intde} by using the new 182 \emph{gold} supernovae Ia (SNe Ia) data \cite{gold182}, the 192 ESSENCE SNe Ia data \cite{essence}, the baryon acoustic oscillation (BAO) measurement from the Sloan Digital Sky Survey (SDSS) \cite{sdss6}, and the shift parameter determined from the three-year Wilkinson Microwave Anisotropy Probe (WMAP3) data \cite{wmap3}. We organized this paper as follows. In Sec. II, we review the basic equations of the interacting HDE model considered in this paper. The choices of the IR cutoff are also discussed in this Sec. II. We fit the model based on the event horizon as the IR cutoff to the observational data above mentioned in Sec III. We end this paper by summarizing our main conclusions in Sec. IV.

\begin{figure*}[t]
\centerline{\psfig{figure=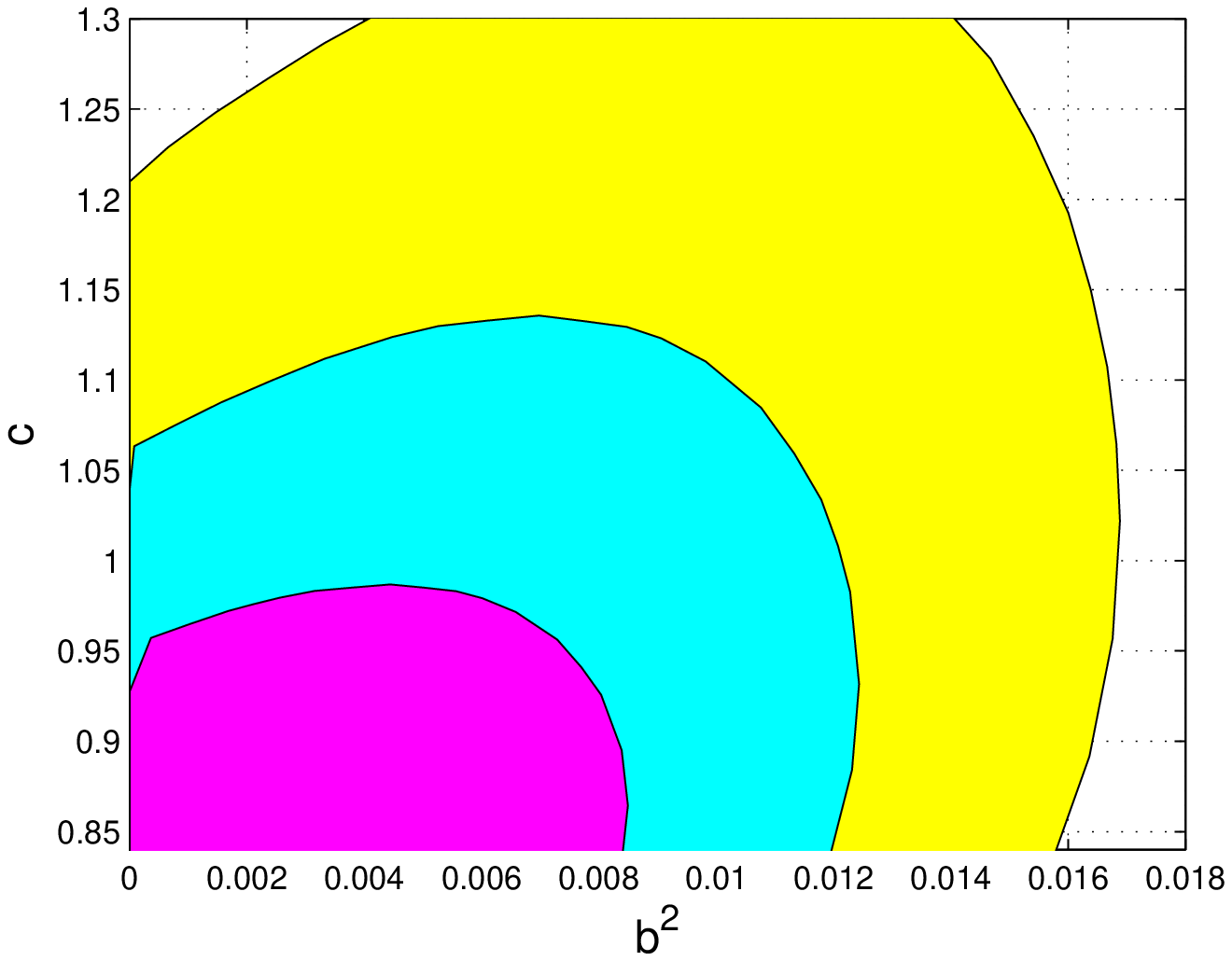,width=3.3truein,height=2.7truein}
\psfig{figure=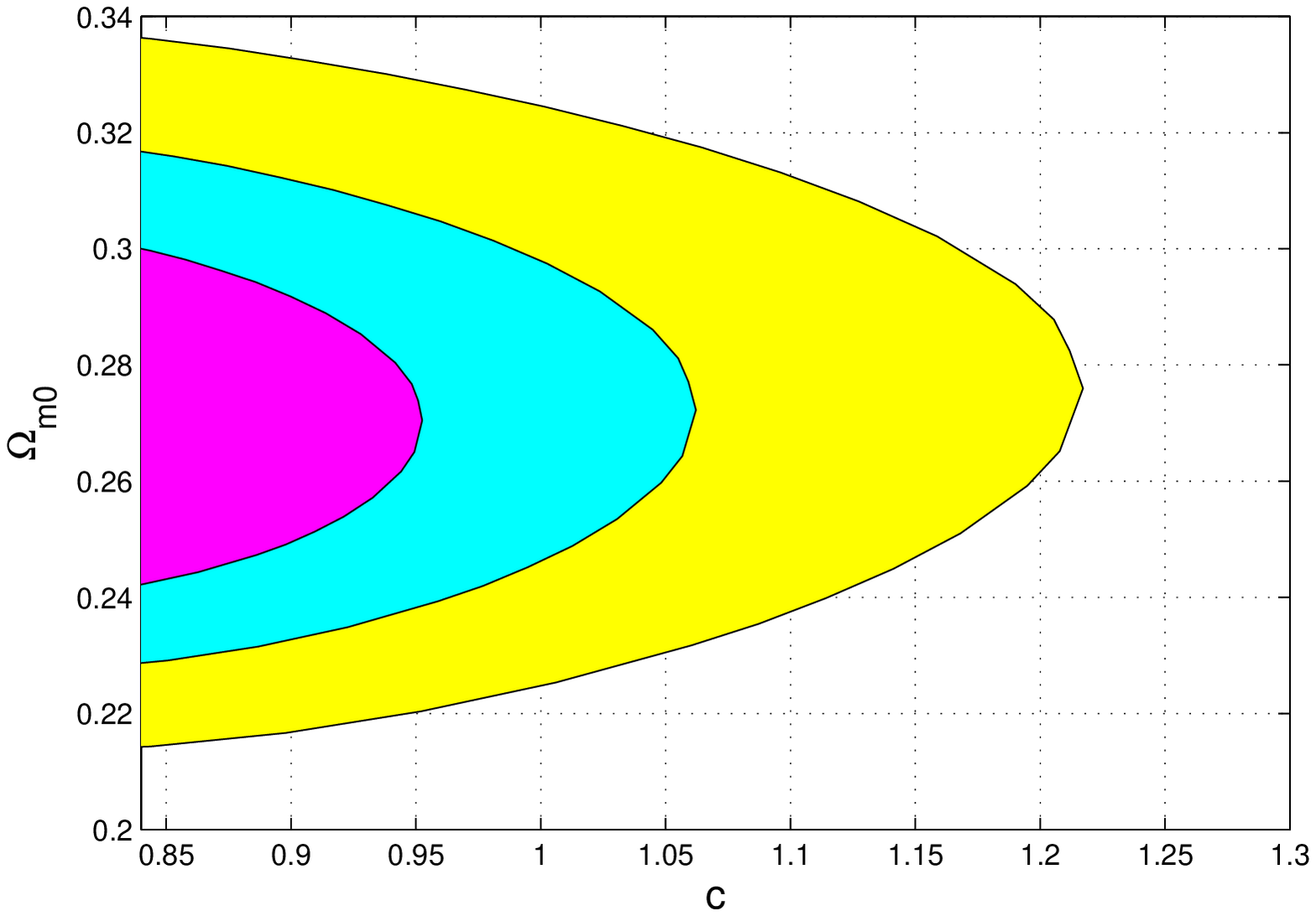,width=3.3truein,height=2.7truein} 
\hskip 0.5in} \caption{The results of our joint analysis involving the ESSENCE (192 SNe Ia) plus BAO plus CMB shift parameter. {\bf{a)}} Confidence contours (1$\sigma$, 2$\sigma$ and 3$\sigma$) in the $b^2 - c$ parametric space. As discussed in the text (see also Table I), at 68.3\% c.l. we find $c=0.85^{+0.18}_{-0.02}$, and $b^2=0.002^{+0.01}_{-0.002}$. {\bf{b)}} Similar results for the $c - \Omega_{m0}$ plane.} 
\label{b2c}
\end{figure*}

\section{Interacting HDE Model}
We consider a spatially flat Friedmann-Robertson-Walker universe with dark matter, HDE and radiation. Due to the interaction between the two dark components, the balance equations between them can be written as
\begin{equation}
\label{meq1}
  \dot{\rho}_m+3H\rho_m = \Gamma \;,
\end{equation}
\begin{equation}
\label{deeq1}
  \dot{\rho}_D+3H(1+\omega_D)\rho_D = -\Gamma
\end{equation}
where the HDE density is
\begin{equation}
\label{rhod}
      \rho_D = 3c^2M_p^2L^{-2}\;.
\end{equation}
In the above equations, $L$ is the IR cutoff, $M_p=1/\sqrt{8\pi G}$ is the reduced Planck mass, $\omega_D$ is the equation of state of the HDE, $\Gamma =9b^2M^2_pH^3$ is a particular interacting term with the coupling constant $b^2$, and the subscript $0$ means the current value of the variable. The HDE, dark matter and radiation density parameters are defined, respectively, as $\Omega_D=\rho_D/(3H^2M^2_p)$, $\Omega_m=\rho_m/(3H^2M^2_p)$, and
$\Omega_\gamma=\rho_\gamma/(3H^2M^2_p)$. Note that, if we choose the Hubble scale as the IR cutoff, i.e., $L=1/H$, then we find that $\Omega_m/\Omega_D=(1-c^2)/c^2$, which means that the HDE always follows the dark matter. Even though the HDE equation of state $w_D$ can be less than $-1/3$ with the help of the interaction \cite{pavon}, this model cannot explain the transition from deceleration to acceleration.

As suggested by Li \cite{Li}, one can choose the future event horizon or particle horizon as the IR cutoff. For the future event horizon,
\begin{equation}
\label{Le}
L(t)=
    a(t)\int_{t}^{\infty}\frac{dt'}{a(t')}=a\int_{a}^{\infty}\frac{da'}{H'a'^2}\;,
\end{equation}
whereas for the particle horizon,
\begin{equation}
\label{lp}
L(t)=a(t)\int_{0}^{t}\frac{dt'}{a(t')}=a\int_{0}^{a}\frac{da'}{H'a'^2}.
\end{equation}
Substituting Eqs. (\ref{Le}) and (\ref{lp}) into Eq. (\ref{rhod}) and taking the derivative
with respect to $x=\ln{a}$, we obtain
\begin{equation}
\label{rhop1}
 \rho'_D\equiv\frac{d\rho_D}{dx}=-6M^2_pH^2\Omega_D\pm\frac{6M^2_p}{c}H^2\Omega^{3/2}_D,
\end{equation}
where the upper (lower) sign is for the event (particle) horizon. Since $\dot{\rho}_D\equiv d\rho_D/dt=\rho'_DH$, Eq. (\ref{deeq1}) can be written as
\begin{equation}
\label{rhop2}
    \rho'_D+3(1+\omega_D)\rho_D=-9M^2_pb^2H^2.
\end{equation}
Combining Eqs. (\ref{rhop1}) and (\ref{rhop2}), we obtain the equation of state of this interacting holographic dark energy, i.e.,
\begin{equation}
\label{omegd}
    \omega_D=-\frac{1}{3}\mp\frac{2}{3}\frac{\sqrt{\Omega_D}}{c}-\frac{b^2}{\Omega_D}.
\end{equation}
When the interaction is absent, $b^2=0$, it is clear from the above expression that we cannot choose the particle horizon as the IR cutoff.
In \cite{kim}, it was argued that the effective equation of state of the HDE in the interacting case
should be
\begin{equation}
\label{wdeff}
\omega_D^{\mathrm{eff}}=\omega_D+\frac{b^2}{\Omega_D}
=-\frac{1}{3}\mp\frac{2}{3}\frac{\sqrt{\Omega_D}}{c}.
\end{equation}
Based on this effective equation of state, it was concluded that there was no phantom crossover even for an interacting HDE model. In fact, by combining the Friedmann equation with Eqs. (\ref{meq1}) and (\ref{deeq1}), we obtain  the acceleration equation
\begin{equation}
\label{acc}
\dot{H}=-4\pi G(\rho+p).
\end{equation}
For a flat universe, the physical consequence of the phantom dark energy is a super-acceleration
when the dark energy dominates. Note that it is $\omega_D$, not $\omega_D^{\mathrm{eff}}$, that
appears in the acceleration equation (\ref{acc}). Therefore, the
effective equation of state seems not to show the true physical meaning of the
equation of state of the HDE, and $\omega_D$ should be used instead.

\begin{figure*}[t]
\centerline{\psfig{figure=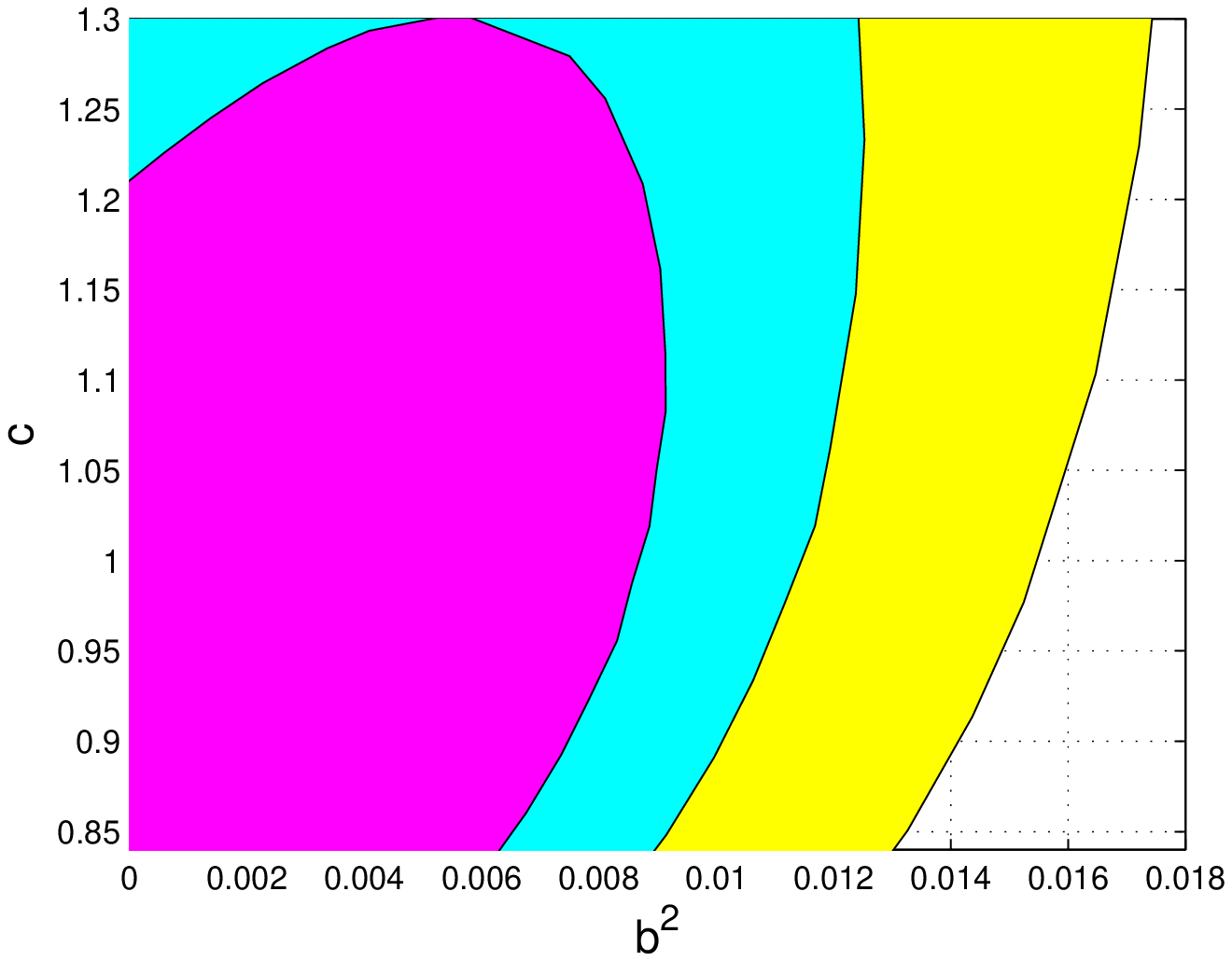,width=3.3truein,height=2.7truein}
\psfig{figure=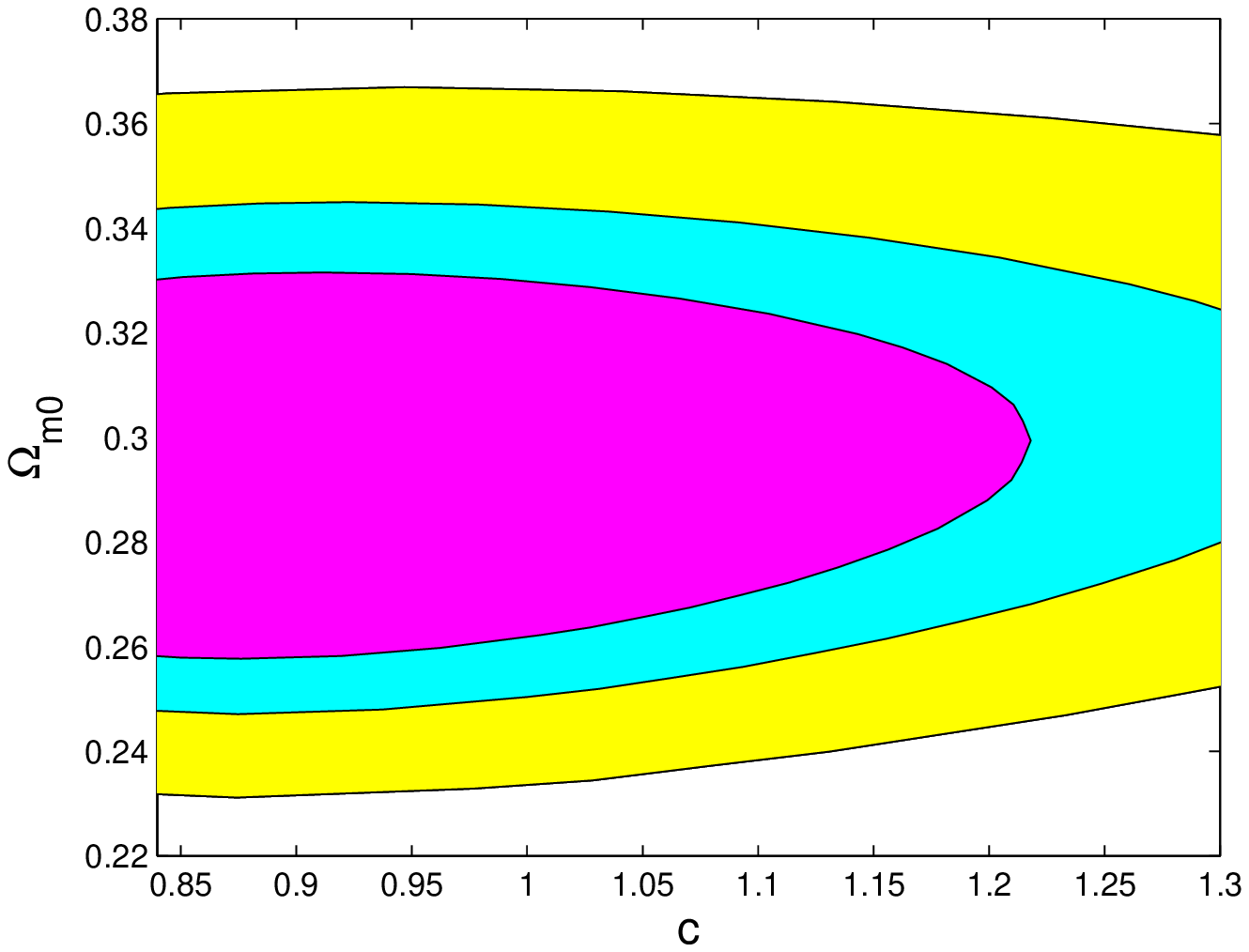,width=3.3truein,height=2.7truein} \hskip
0.5in} \caption{{\bf{a)}} The same as in Figure 1 when the ESSENCE (192 SNe Ia) data are replaced by the new 182 \emph{Gold} sample. The contours in the $b^2 - c$ plane also correspond to $1\sigma$, $2\sigma$ and $3\sigma$. {\bf{b)}} The same as in the previous Panel for the $c - \Omega_{m0}$ parametric plane.} 
\label{goldb2c}
\end{figure*}

Substituting Eq. (\ref{omegd}) into Eq. (\ref{rhop2}) and applying the
definition of $\Omega_D$, we have
\begin{equation}\label{hp1}
    \frac{H'}{H}=-\frac{\Omega'_D}{2\Omega_D}-1\pm\frac{\sqrt{\Omega_D}}{c}.
\end{equation}
On the other hand, substituting
$\dot{H}=H'H$ and $p_D=\omega_D\rho_D$ into Eq. (\ref{acc}), we obtain
\begin{equation}\label{hp2}
    \frac{H'}{H}=\frac{1}{2}\Omega_D\pm\frac{\Omega^{3/2}_D}{c}+\frac{3}{2}b^2-\frac{3}{2}-\frac{1}{2}\Omega_\gamma.
\end{equation}
Now, combining Eqs. (\ref{hp1}) and (\ref{hp2}), we find the
differential equation for $\Omega_D$, i.e.,
\begin{equation}\label{omdp}
  \frac{\Omega_D'}{\Omega_D} =
  1-\Omega_D\pm\frac{2\sqrt{\Omega_D}}{c}(1-\Omega_D)-3b^2+\Omega_\gamma\;,
\end{equation}
a result that is consistent with Eq. (5) of Ref.~\cite{bin05}
when the radiation term $\Omega_\gamma$ is neglected.

If we choose the particle horizon as the IR cutoff, the current acceleration requires that $\omega_{D0}<-1/3-(\Omega_{m0}+2\Omega_{\gamma0})/3\Omega_{D0}$. From Eq. (\ref{omegd}), we also obtain a lower bound on $b^2$,
\begin{equation}
\label{b2gt}
 b^2>\frac{\Omega_{m0}}{3}+\frac{2}{3}\frac{\Omega_{D0}^{3/2}}{c}+\frac{2}{3}\Omega_{\gamma0}.
\end{equation}
The past deceleration and the transition from deceleration to acceleration requires that
$\Omega_{D0}'\ge 0$, so Eq. (\ref{omdp}) gives the upper bound on $b^2$,
\begin{equation}
\label{b2lt}
b^2\le\frac{\Omega_{m0}}{3}\left(1 - 2\frac{ \sqrt{\Omega_{D0}}}{c}\right)+\frac{1}{3}\Omega_{\gamma0}.
\end{equation}
By comparing Eqs. (\ref{b2gt}) and (\ref{b2lt}), we see that the upper bound is lower than
the lower bound, so that the inequalities are not satisfied. The model based on the particle
horizon as the IR cutoff is not, therefore, a viable dark energy model. In what follows, we consider only
the HDE based on the event horizon. As discussed in
\cite{bin05}, the interaction $\Gamma$ cannot be too strong and the parameters
$b^2$ and $c$ are not totally free; they need to satisfy some constraints.
Following \cite{bin05}, we take $0\le b^2\le 0.2$ and $\sqrt{\Omega_D}<c<1.255$.

\begin{figure*}[t]
\centerline{\psfig{figure=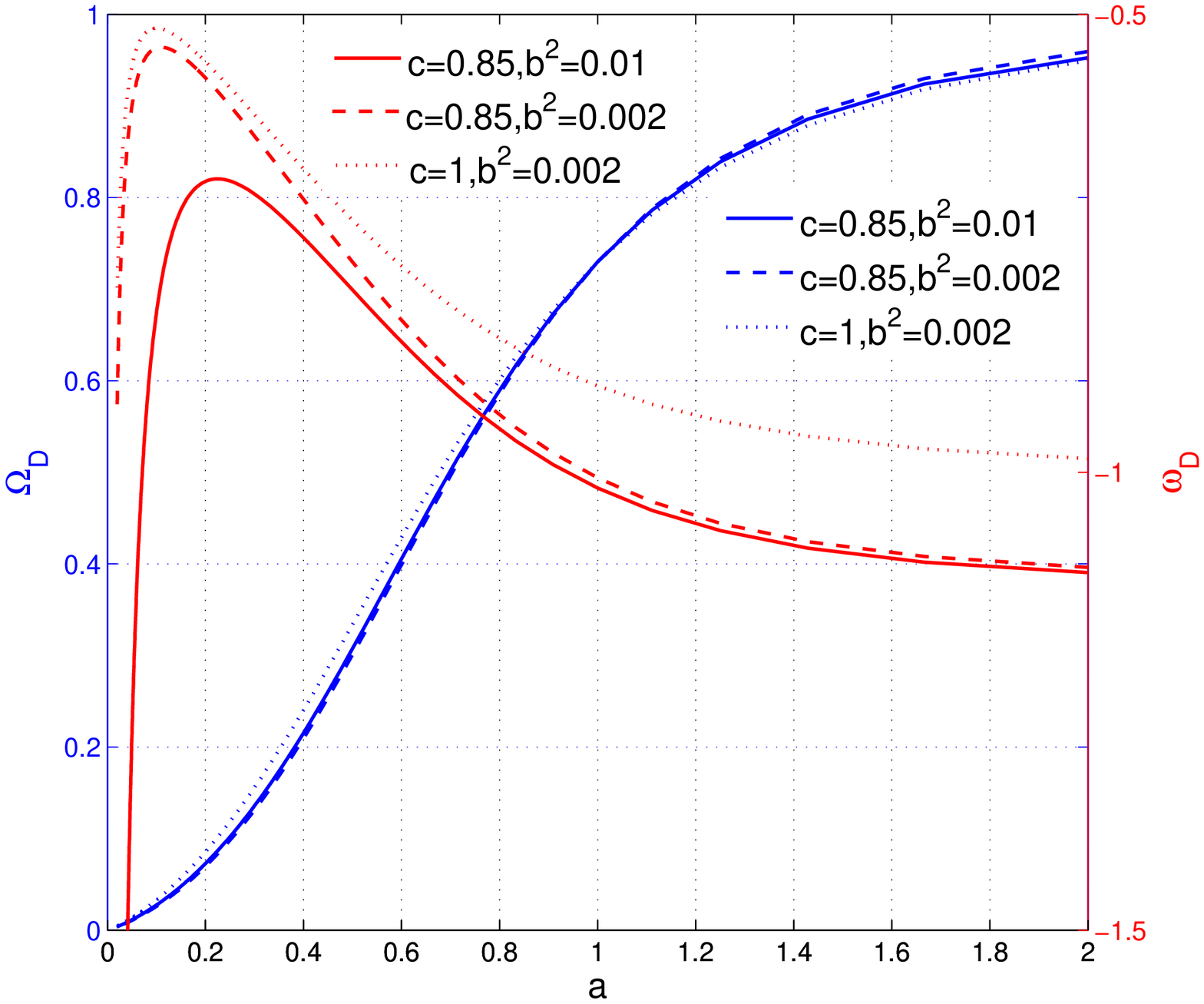,width=3.3truein,height=2.7truein}
\psfig{figure=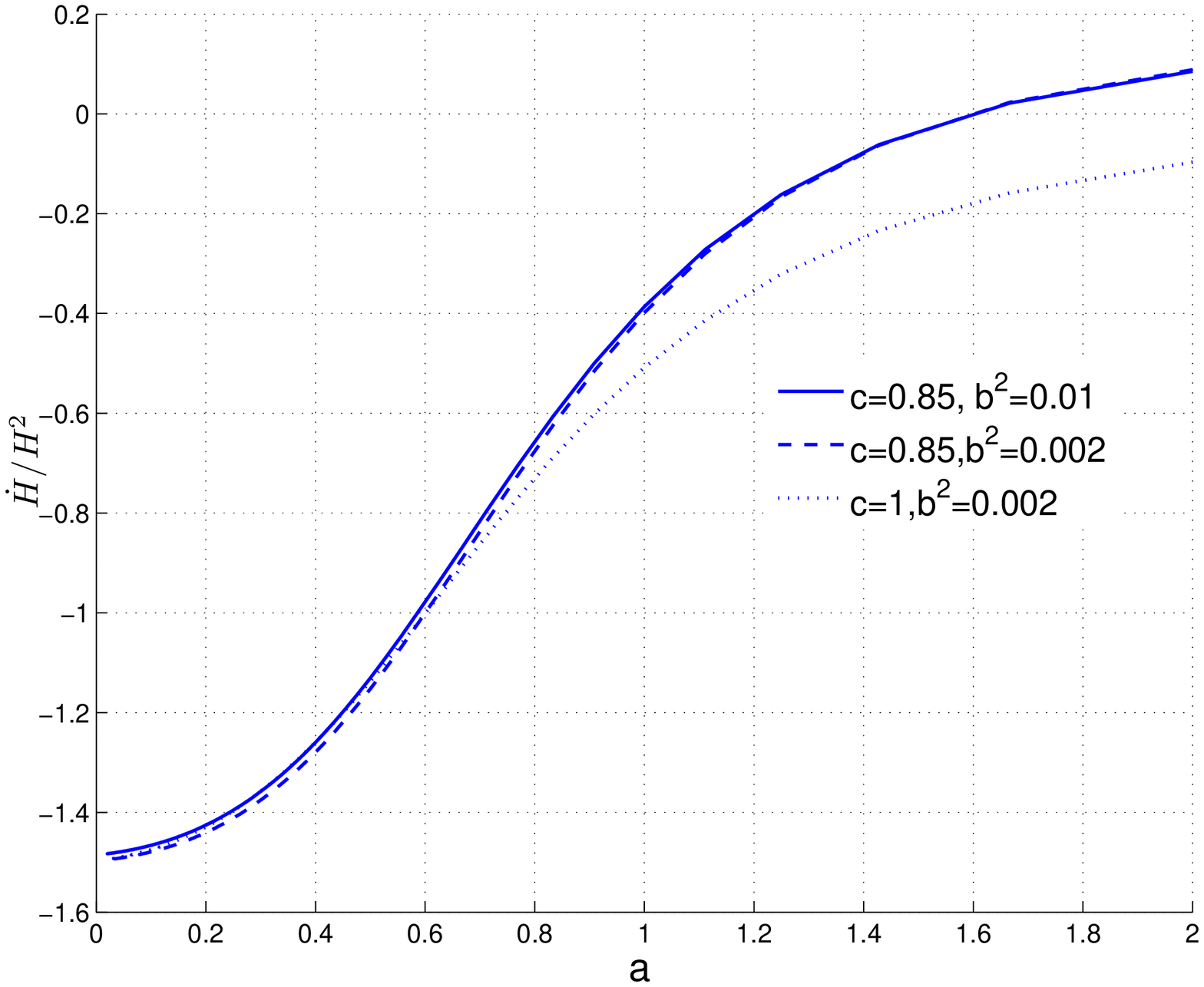,width=3.3truein,height=2.7truein}
\hskip 0.5in}
\caption{{\bf{a)}} Evolution of $\Omega_D$ and $\omega_D$ with the scale factor $a$. To plot these curves we have fixed the best-fit value of $\Omega_{m0}=0.27$. The solid, dashed and dotted lines stand, respectively, for the pairs ($b^2=0.01$, $c=0.85$), ($b^2=0.002$, $c=0.85$), and ($b^2=0.002$, $c=1$). {\bf{b)}} Evolution of $\dot{H}/H^2$ with the scale factor $a$. Note that, as $\omega_D$ is becoming more and more negative and crosses the phantom divide line (Panel 3a), the function $\dot{H}$ increases from negative to positive values. As in the previous Panel, the value of the matter density parameter has been fixed at $\Omega_{m0}=0.27$ and the solid, dashed and dotted lines correspond to the above combinations of the parameters $b^2$ and $c$.}
\label{3line}
\end{figure*}

\section{Observational Constraints}

\subsection{The data}

There are three parameters $\Omega_{m0}$, $c$ and $b^2$ in the interacting HDE model since $\Omega_{D0}=1-\Omega_{m0}-\Omega_{\gamma 0}$
and $\Omega_{\gamma0} \sim 10^{-5}$. In order to place limits on them and test the viability of the model, we apply both the 182 gold SNe Ia \cite{gold182} and the 192 ESSENCE SNe Ia data \cite{essence} to fit these parameters by minimizing
\begin{equation}
\label{chi}
\chi^2 = \sum_{\substack{i}}\frac{[\mu_{obs}(z_i)-\mu(z_i)]^2}{\sigma^2_i},
\end{equation}
where the extinction-corrected distance modulus $\mu(z)=5\log_{10}(d_L(z)/\mathrm{Mpc})+25$, $\sigma_i$ is the total uncertainty in the $\mu_{obs}$ observations, and the luminosity distance is given by
\begin{eqnarray}
    d_L &=& (1+z)\int_{0}^{z}\frac{\mathrm{d}z'}{H(z')}\nonumber\\
        &=&[\frac{c(1+z)^2}{H(z)\sqrt{\Omega_D(z)}}-(1+z)\frac{c}{\sqrt{\Omega_{D0}}H_0}],
\end{eqnarray}
where $z=a_0/a-1$. In all the subsequent analyses, we have marginalized the Hubble parameter $H_0$.

In addition to the SNe Ia data, we also use the BAO measurement from the SDSS data \cite{sdss6,wmap3}
\begin{equation}
\label{sdss}
\begin{split}
 A&=\frac{\sqrt{\Omega_{m0}}}{E(0.35)^{1/3}}\left[\frac{1}{0.35}\int_{0}^{0.35}\frac{dz}{E(z)}\right]^{2/3}\\
 &=0.469\left(\frac{0.95}{0.98}\right)^{-0.35}\pm 0.017,
 \end{split}
\end{equation}
and the CMB shift parameter measured from WMAP3 data \cite{yun06,wmap3}
\begin{equation}
\label{cmb}
\mathcal{R}=\sqrt{\Omega_{m0}}\int_{0}^{z_{ls}}\frac{dz}{E(z)}=1.70\pm 0.03,
\end{equation}
where the dimensionless function $E(z) = H(z)/H_0$ and $z_{ls}=1089\pm 1$.
In order to obtain the distance, we need to find out the evolution of $\Omega_D(z)$
and $H(z)$, so we need to solve Eqs. (\ref{hp2}) and (\ref{omdp})
numerically. Since the derivatives in Eqs. (\ref{hp2}) and (\ref{omdp}) are
with respect to $x=\ln a$, we need to rewrite them with respect to $z$. We find
\begin{equation}
\label{hpz}
\frac{d H}{dz}=-\frac{H}{1+z}\left(\frac{1}{2}\Omega_D +\frac{\Omega^{3/2}_D}{c}
+\frac{3}{2}b^2-\frac{3}{2}-\frac{1}{2}\Omega_\gamma\right),
\end{equation}
and
\begin{equation}
\label{omdpz}
\frac{d\Omega_D}{dz} =-\frac{\Omega_D}{1+z}\left[
(1-\Omega_D)(1 + \frac{2\sqrt{\Omega_D}}{c})-3b^2+\Omega_\gamma\right].
\end{equation}
By solving numerically the above equations, we then obtain the evolutions of $\Omega_D$ and $H$ as a function of the redshift.

\begin{table*}
\begin{center}
\caption{The best-fit results for the HDE parameters}
\label{sntab}
\begin{tabular}{|c|c|c|c|c|c|}
\hline
Model&Results &Gold   &  Gold+$A$+$\mathcal{R}$ &ESSENCE &ESSENCE+$A$+$\mathcal{R}$ \\
\hline
  &$\chi^2$ &158.27    & 158.66      &195.34        & 196.16  \\
With  &$\Omega_{m0}$  & $0.32^{+0.29}_{-0.13}$  & $0.29\pm 0.04$  &$0.27^{+0.23}_{-0.15}$ & $0.27^{+0.04}_{-0.03}$ \\
Interaction  &$b^2$  &$0^{+0.2}_{-0}$  & $0^{+0.01}_{-0}$  &$0.02^{+0.09}_{-0.02}$   & $0.002^{+0.01}_{-0.002}$  \\
&$c$  &$0.82^{+0.48}_{-0.18}$  & $0.88^{+0.40}_{-0.07}$ &$0.85^{+0.45}_{-0.18}$  & $0.85^{+0.18}_{-0.02}$  \\
\hline
&$\chi^2$& 158.27 &158.66 & 195.75 & 196.29 \\
$b^2=0$&$\Omega_{m0}$& $0.31^{+0.07}_{-0.1}$ & $0.29\pm 0.03$ & $0.27^{+0.03}_{-0.14}$ & $0.27^{+0.03}_{-0.02}$ \\
&$c$& $0.82^{+0.48}_{-0.04}$ & $c=0.88^{+0.24}_{-0.06}$ & $c=0.85^{+0.45}_{-0.02}$ & $0.85^{+0.1}_{-0.02}$ \\
\hline
$\Lambda$CDM & $\chi^2$& 158.49 & 161.87 & 195.34 & 196.12 \\
& $\Omega_{m0}$ & $0.34\pm 0.04$ & $0.29\pm 0.02$ & $0.27\pm 0.03$ & $0.27\pm 0.02$\\
\hline
\end{tabular}
\end{center}
\end{table*}

\subsection{Results}

In Figs. (1) and (2) we show the results of our statistical analyses. Figure (1a) shows the $c - b^2$ plane for the joinf analysis involving the 192 ESSENCE SNe Ia data \cite{essence} and the other cosmological observables discussed above. For this analysis the best fit values are $\Omega_{m0}=0.27^{+0.04}_{-0.03}$,  $c=0.85^{+0.18}_{-0.02}$, and $b^2=0.002^{+0.01}_{-0.002}$ (at 68.3\% c.l.) with $\chi_{min}^2=196.16$. If we fix $c=1$, we find $\Omega_{m0}=0.26^{+0.04}_{-0.03}$ and $b^2=0.005^{+0.008}_{-0.005}$ (at 68.3\% c.l.) with $\chi^2=198.96$. The plane $\Omega_{m0} - c$ is shown in Panel (1b). Figure (2a) shows the same parametric space $c - b^2$ when the 192 ESSENCE data is replaced by the new 182 \emph{gold} sample \cite{gold182}. In this case, we find $\Omega_{m0}=0.29\pm 0.04$, $b^2=0^{+0.01}_{-0}$, $c=0.88^{+0.40}_{-0.07}$ (at 68.3\% c.l.) with $\chi^2= 158.66$. The plane $\Omega_{m0} - c$ for this latter combination of data is also shown in Panel (2b). We note that from both combinations the value of $b^2$ is very close to 0, which suggests a very weak coupling or a noninteracting HDE. Such a result is also in agreement with the limits found in Ref.~\cite{intde}. By fixing $b^2=0$, we also fit the HDE model without interaction to the observational data discussed above. These results are summarized in Table \ref{sntab}. For the sake of comparison, we also list the best-fit results for a flat $\Lambda$CDM model.

Finally, by fixing the value of the matter density parameter at $\Omega_{m0}=0.27$ we show, in Fig. (3a), the evolutions of $\Omega_D$ and $\omega_D$ with the scale factor $a$. The behavior of $\dot{H}/H$ is also shown in Fig. (3b). The curves displayed in these Panels are complementary in the sense that from them, we see that while $\omega_D$ crosses the cosmological constant barrier to the phantom region, $\dot{H}$ increases from negative to positive values. The distinctive future super-acceleration, which is an evidence of a phantom behavior, is apparent from Panel (3b).

\section{Conclusions}
We have analyzed the observational viability of an interacting HDE model characterized by the parameters $c$ and $b^2$ and shown that in order to make it compatible with current observational data only the future event horizon can be used as the IR cutoff. By assuming a particular interacting term and fitting the model to the observational data, we have found that an HDE is capable of explaining the current observations of SNe Ia, BAO and shift parameter from CMB data for the intervals of the parameters $\Omega_{m0}$, $c$ and $b^2$ displayed in Table I. From both analyses (involving the 192 SNe Ia ESSENCE data and the new 182 \emph{Gold} sample) we have also found that a very weak coupling or even a noninteracting HDE model is favored ($b^2 \simeq 0$). By using the fitting results for the parameters, we have shown that the expansion rate in the interacting HDE model can be super-accelerated, i.e. $\dot{H}>0$, which is the key physical consequence of the so-called phantom behavior \cite{phantom}. This phantom regime is obtained after a transition from $\omega_D>-1$ to $\omega_D<-1$ (phantom crossing behavior), which is a feature well studied in the context of some non-minimally coupled scalar fields \cite{peri}.

\begin{acknowledgments}
J.S. Alcaniz would like to thank the hospitality of the Physics Department of Baylor University where this work was developed. Y.G. Gong is supported by Baylor University, NNSFC under grant No. 10447008 and 10605042, and SRF for ROCS, State Education Ministry. A. Wang is partially supported by VPR funds,  Baylor University. J.S. Alcaniz is supported by FAPERJ No. E-26/171.251/2004 and CNPq No. 307860/2004-3.

\end{acknowledgments}

\end{document}